# Optimal rate-variance coding due to firing threshold adaptation near criticality

Mauricio Girardi-Schappo[1,2,*], Leonard Maler[3], André Longtin[2,3]

[1] Departamento de Física - Universidade Federal de Santa Catarina - Florianópolis SC - 88040-900 – Brazil
[2] Department of Physics - University of Ottawa - Ottawa ON - K1N 6N5 - Canada
[3] Department of Cellular and Molecular Medicine - University of Ottawa - Ottawa ON - K1H 8M5 – Canada
* Corresponding author: girardi.s@gmail.com

4 September 2025

**Abstract**

Recurrently connected neuron populations play key roles in sensory perception and memory storage across various brain regions. While these populations are often assumed to encode information through firing rates, this method becomes unreliable with weak stimuli. We propose that in such cases, information can be transmitted via spatial spike patterns, employing a sparse or combinatorial coding based on firing rate variance. Around the critical point of a stochastic recurrent excitable network, we uncover a synergistic dual-coding scheme, enabled by single-cell threshold adaptation. This scheme optimizes variance coding for weak signals without compromising rate coding for stronger inputs, thus maximizing input/output mutual information. These optimizations are robust across adaptation rules and coupling strengths through self-suppression of internal noise, particularly around the network's phase transition, and are linked to threshold recovery times observed in hippocampal memory circuits (~$10^2$–$10^3$ms). In contrast, nonadaptive networks perform similarly only at criticality, suggesting that threshold adaptation is essential for reliable encoding of weak signals into diverse spatial patterns. Our results imply a fundamental role for near-critical latent adaptive dynamics enabled by dual coding in biological and artificial neural networks.

The nonlinear statistical dynamics that support the encoding of information by the brain remain elusive. Candidate physical theories consider the representation of inputs in terms of firing rates or spike patterns across neuronal populations. Optimal rate coding (RC) capacity (Kinouchi & Copelli, 2006; Shew *et al.*, 2009) and information processing (Beggs & Plenz, 2003; Shew *et al.*, 2011) have been associated to branching process critical point dynamics, a framework shown to share a universal scaling law with actual brain activity (Carvalho *et al.*, 2021). However, firing rates barely respond to weak signal inputs and cannot account for the behavioral performance of neural systems near threshold perception (Fettiplace & Hackney, 2006; Gussin *et al.*, 2007; Jacobs *et al.*, 2009; Heil & Peterson, 2015; Jung *et al.*, 2016). As an alternative, spike patterns have been implicated in auditory (Feigin *et al.*, 2021), olfactory (Saha *et al.*, 2015), visual (Osborne *et al.*, 2008) and electrosensory (Nesse *et al.*, 2021) perception as well as higher brain function (Cayco-Gajic *et al.*, 2017), generating sparse and/or combinatorial codes – both being examples of *pattern coding* (PC) where the precise constellation of activated cells matters. PC increases memory capacity, allows robust encoding of external signals, facilitates the transmission of information, and saves energy (Olshausen & Field, 2004).

Can neural systems synergistically explore both strategies to encode information? Here we show that they can if the ubiquitous property of adaptation (Benda, 2021) is part of the dynamics. Spiking frequency can slow down by raising the firing threshold after each spike. This keeps track of time (Itskov *et al.*, 2011), improves information transmission (Chacron *et al.*, 2007), and enhances feature selectivity (Wilent & Contreras, 2005), coding precision (Huang *et al.*, 2016) and synchrony detection (Azouz & Gray, 2003). Slowly recovering thresholds have been reported in mammalian cortex (Fleidervish & Gutnick, 1996), hippocampus (Mickus *et al.*, 1999; Ellerkmann *et al.*, 2001; Trinh *et al.*, 2023) and an analogous teleost pallial region (Trinh *et al.*, 2019). These areas are fundamental for memory tasks (Pastalkova *et al.*, 2008) and discriminating similar sensory inputs (Barak *et al.*, 2013) and would benefit from optimal PC. Adaptive thresholds combined with other mechanisms are known to support the self-organization of networks toward a near-critical regime (Menesse *et al.*, 2022), but their isolated effect on coding near the critical point was never explored. We investigated this effect and unveiled a parametrically robust mechanism of optimal coding due to threshold adaptation near an absorbing critical point.

PC provides an optimal representation of stimuli that are too weak to be discriminable by RC, extending the system's coding capacity. In a recurrent excitable network, we show that this dual code is robustly accomplished by networks of neurons whose individual thresholds quickly increase at each spike and slowly decay between spikes. Conversely, networks of nonadaptive neurons perform similarly only in a narrow near-critical regime. We develop a theory for coding in near-critical adaptive systems, emphasizing its differences with the current feedback theories for self-organization near a critical point (Buendía *et al.*, 2020; Kinouchi *et al.*, 2020). The threshold adaptation timescales of hippocampus neurons (Trinh *et al.*, 2023) are within the range where we found coding optimization. This makes threshold adaptation a key component of coding, and suggests that PC improves the discrimination between similar inputs, possibly enhancing diverse cognitive tasks and computational applications, such as artificial intelligence.

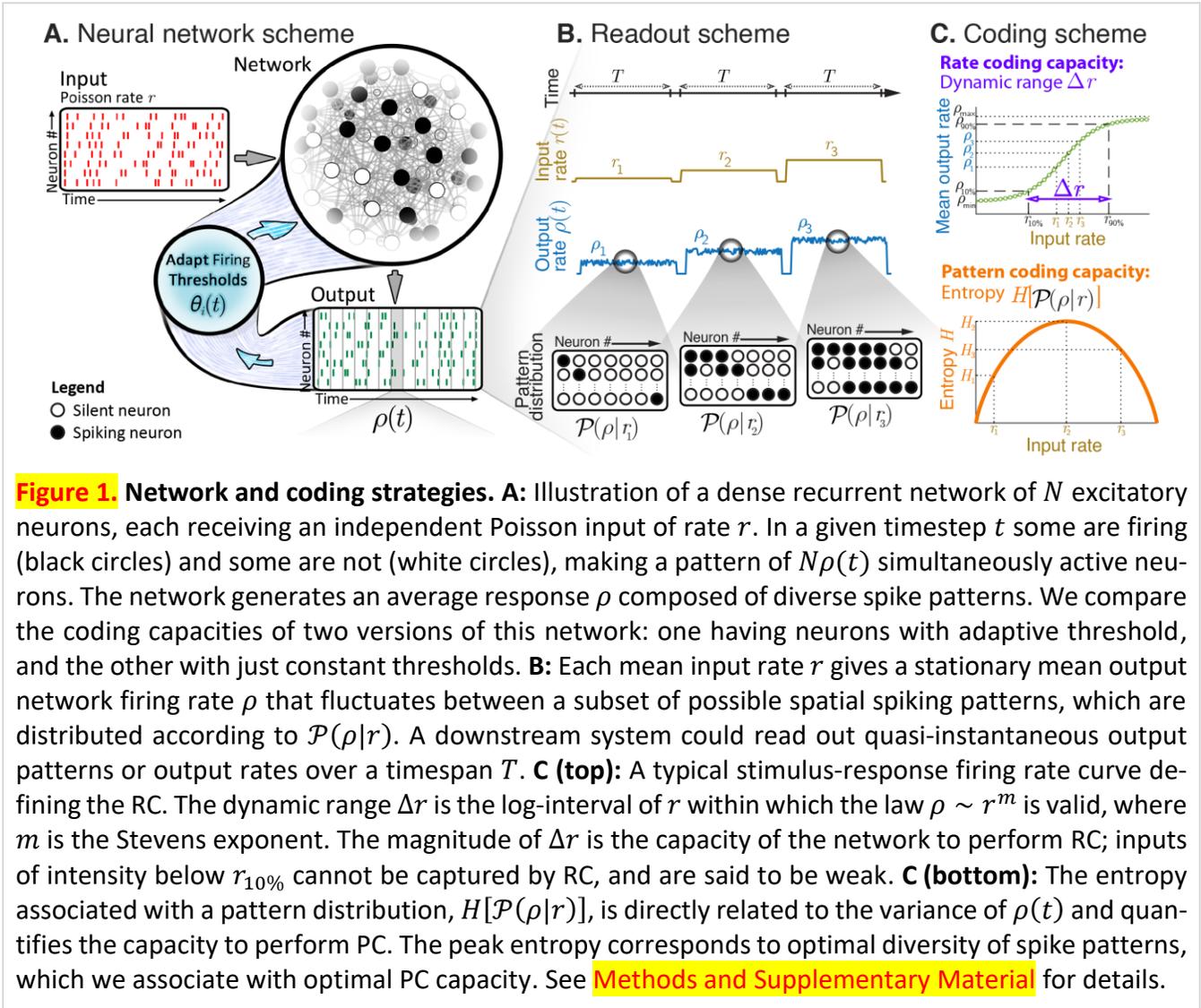

**Figure 1. Network and coding strategies. A:** Illustration of a dense recurrent network of $N$ excitatory neurons, each receiving an independent Poisson input of rate $r$. In a given timestep $t$ some are firing (black circles) and some are not (white circles), making a pattern of $N\rho(t)$ simultaneously active neurons. The network generates an average response $\rho$ composed of diverse spike patterns. We compare the coding capacities of two versions of this network: one having neurons with adaptive threshold, and the other with just constant thresholds. **B:** Each mean input rate $r$ gives a stationary mean output network firing rate $\rho$ that fluctuates between a subset of possible spatial spiking patterns, which are distributed according to $\mathcal{P}(\rho|r)$. A downstream system could read out quasi-instantaneous output patterns or output rates over a timespan $T$. **C (top):** A typical stimulus-response firing rate curve defining the RC. The dynamic range $\Delta r$ is the log-interval of $r$ within which the law $\rho \sim r^m$ is valid, where $m$ is the Stevens exponent. The magnitude of $\Delta r$ is the capacity of the network to perform RC; inputs of intensity below $r_{10\%}$ cannot be captured by RC, and are said to be weak. **C (bottom):** The entropy associated with a pattern distribution, $H[\mathcal{P}(\rho|r)]$, is directly related to the variance of $\rho(t)$ and quantifies the capacity to perform PC. The peak entropy corresponds to optimal diversity of spike patterns, which we associate with optimal PC capacity. See Methods and Supplementary Material for details.

## *Results*

We study a dense excitable network of $N$ stochastic excitatory integrate-and-fire neurons with average synaptic coupling $J$ (Figure 1; see Methods and Supplementary Material for the definition of the model, measurements, calculations and extra figures). Random input patterns are injected in the network by independent Poisson processes of mean rate $r$ acting on each neuron, and the network produces a stationarily fluctuating output firing rate $\rho$. This input increases the probability that the neuron fires in any given time step. The dynamic range $\Delta r$ measures the capacity to perform RC. It represents the extent of inputs where the stimulus-response curve follows $\rho \sim r^m$ ($m$ is the Stevens exponent).

We compute the distribution $\mathcal{P}(\rho|r)$ of observed outputs given fixed input rates to each neuron. This yields the conditional entropy $H[\mathcal{P}(\rho|r)]$ as the capacity of the network to encode stimuli into instantaneous spike patterns, since $\mathcal{P}(\rho|r)$ is a measure of the multiplicity of neuron combinations that fire in response to an input. Thus, interpreting the entropy from the Boltzmann-Gibbs point of view, maximal conditional entropy implies in maximal capacity for generating multiple patterns given an input.

We know from Gaussian signals that both the entropy and the variance of $\rho$ are related (Warland *et al.*, 1996). Moreover, in a MF network, a pattern is completely determined by $\rho$ (the fraction of active neurons in the network due to the all-to-all connectivity that breaks down the notion of spatial position). This allows us to use the variance of $\rho$ to quantify PC. We are interested in studying whether the $\text{Var}(\rho)$ can be used to encode inputs when stimuli are too weak to be captured by the mean of $\rho$ (*i.e.*, by RC). Optimal PC, then, requires variance-encoding together with maximal entropy in the regime of weak inputs.

Switching to the Shannon interpretation of entropy, we measure the information drop in the output due to the input $I[\rho; r]$, and refer to it as mutual information (MI – see Methods). We employ this metric to quantify the processing of information (Beggs & Plenz, 2003; Osborne *et al.*, 2008; Shew *et al.*, 2011).

We compare these measurements for networks where all neurons have either constant (Figure 2) or adaptive thresholds (Figure 3). We tested three different adaptation rules and found the same qualitative results for all of them (see Supplementary Material). The constant threshold network presents a mean-field directed percolation (MF-DP) critical point at a specific value of the average synaptic coupling strength, namely, $J_c = 5$ (Figure S1). Hence, we use the coupling strength to probe for the behavior of both versions of the network around this critical point.

### A. Constant threshold networks cannot robustly encode weak inputs

Constant threshold networks perform optimal RC at the critical point (Figures 2A-B), since the Stevens exponent becomes the critical exponent associated with the external field (Kinouchi & Copelli, 2006), $m = 1/\delta_h = 1/2$, Figure 2C. These results are in agreement with other models and experiments (Shew *et al.*, 2009; Shew *et al.*, 2011). The MI is maximized at the critical point. However, the optimizations of either the dynamic range or the MI are not robust, since they rely on being at or near the critical point, which is a fine-tuning of synaptic strengths $J \sim J_c$, and characterizes a poor biological strategy.

Our system is known to possess a MF-DP critical point, so it displays the standard power law (PL) avalanche distributions with the expected scaling when $r \to 0$. We studied the distortion of avalanche distributions with respect to the input, and observed that the PL regime shrinks as $r$ grows (Figure S2).

Inputs smaller than $r_{10\%}$ (a standard lower bound of the dynamic range; see Methods) cannot produce noticeable changes in the average output, and hence are considered weak. That is, $r = 10^{-6}$ spikes/ms/neuron = 1 spike every 10 ms in the population, is used as a reference for weak stimuli. These stimuli could still be transduced by means of PC. However, the entropy (Figures 2A-B) and variance (Henkel *et al.*, 2008) remain constant for the supercritical regime, preventing the encoding of patterns into $\text{Var}(\rho)$ (Figure 2C).

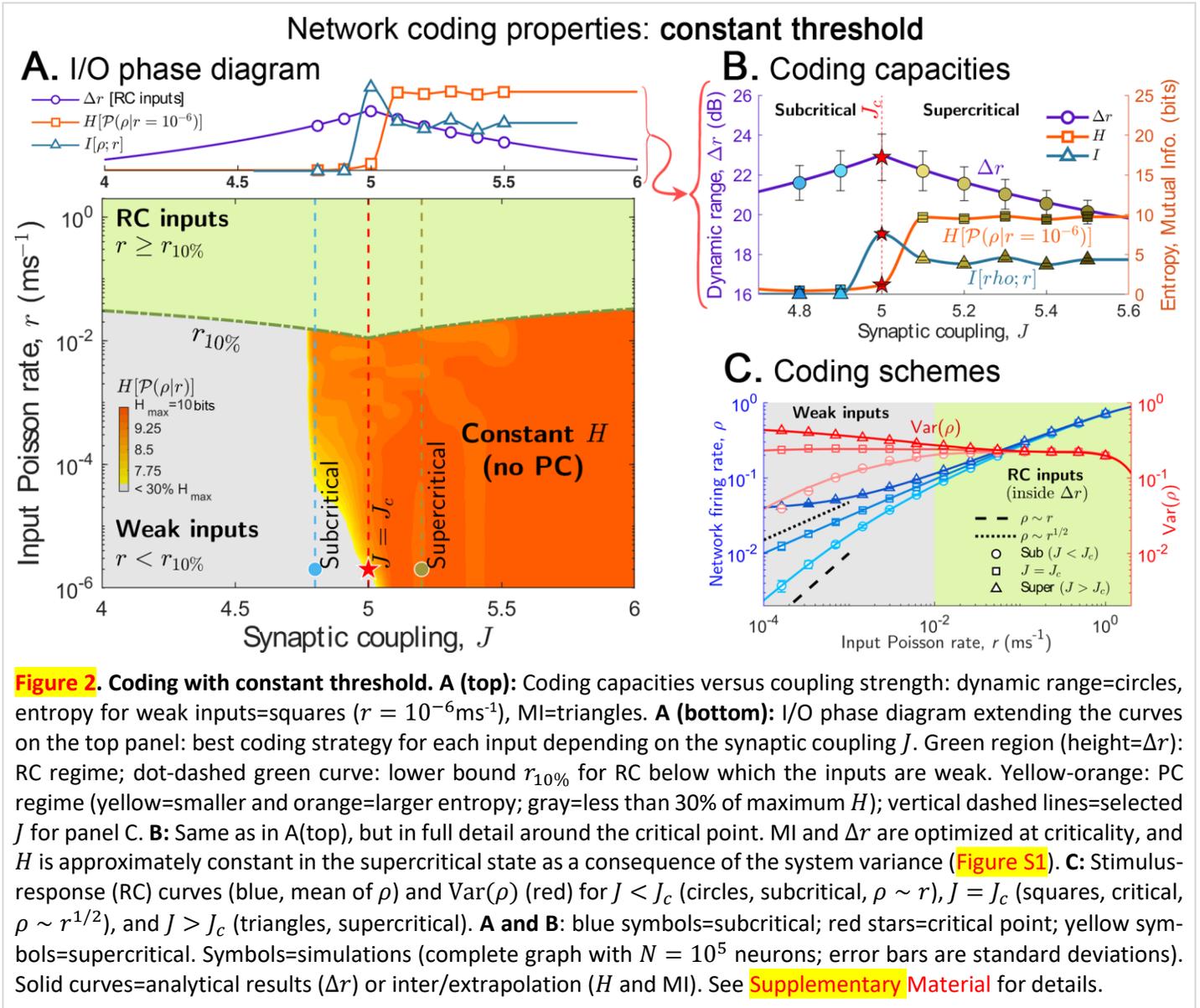

**Figure 2. Coding with constant threshold. A (top):** Coding capacities versus coupling strength: dynamic range=circles, entropy for weak inputs=squares ($r = 10^{-6}$ ms$^{-1}$), MI=triangles. **A (bottom):** I/O phase diagram extending the curves on the top panel: best coding strategy for each input depending on the synaptic coupling $J$. Green region (height=$\Delta r$): RC regime; dot-dashed green curve: lower bound $r_{10\%}$ for RC below which the inputs are weak. Yellow-orange: PC regime (yellow=smaller and orange=larger entropy; gray=less than 30% of maximum $H$); vertical dashed lines=selected $J$ for panel C. **B:** Same as in A(top), but in full detail around the critical point. MI and $\Delta r$ are optimized at criticality, and $H$ is approximately constant in the supercritical state as a consequence of the system variance (Figure S1). **C:** Stimulus-response (RC) curves (blue, mean of $\rho$) and Var($\rho$) (red) for $J < J_c$ (circles, subcritical, $\rho \sim r$), $J = J_c$ (squares, critical, $\rho \sim r^{1/2}$), and $J > J_c$ (triangles, supercritical). **A and B**: blue symbols=subcritical; red stars=critical point; yellow symbols=supercritical. Symbols=simulations (complete graph with $N = 10^5$ neurons; error bars are standard deviations). Solid curves=analytical results ($\Delta r$) or inter/extrapolation ($H$ and MI). See Supplementary Material for details.

### B. Threshold adaptation yields good and robust RC for large timescales

We use the recovery timescale $\tau$ of the threshold as a control parameter to trace the I/O phase diagram (Figure 3A-B). This quantity was recently observed to be in the range of 100ms to 1,000ms in both CA3 neurons and mossy cells in the hilus (Trinh et al., 2023). The slope of the stimulus-response curve is $m = 1$ (Figure 3C), making the RC suboptimal in comparison with a constant threshold system. However, the RC capacity, measured by the size of the dynamic range, asymptotically grows with increasing $\tau$, reaching $\Delta r \sim 14 - \mathcal{O}(1/\tau)$ (Figures 3A-B). These findings hold for subcritical, critical and supercritical synaptic couplings $J$ (see as $\rho$ is constant as a function of $J$ in Figure S4A). These results are reproduced for alternative formulations of the threshold dynamics (Figures S8,S9).

However, adaptive networks converge to a stable firing rate $\rho \sim 1/\tau$ for weak inputs due to near critical fluctuations generated by the adaptation (Figures 3C). Since this regime does not change with $r$, weak inputs cannot be encoded in the mean firing rate of adaptive networks.

*C. Threshold adaptation causes optimal and robust PC and MI*

The variance of the firing rate can encode weak inputs (Figures 3C). Notice that even though the firing rate is flat at $\rho \sim 1/\tau$, $\text{Var}(\rho)$ increases with decreasing $r$, encoding weak stimuli and enabling PC. This phenomenon was observed throughout the highlighted area of the I/O phase diagram, from $\tau = 100$ms to $\tau = 10,000$ms. Although surprising, this can be explained by the fact that threshold adaptation acts to filter out high firing rate activities, allowing only slower fluctuations to make up $\rho \sim 1/\tau$. These become critical in the limit of $\tau \to \infty$ and $r \to 0$, making $\text{Var}(\rho)$ (*i.e.*, the network susceptibility) grow as $r$ vanishes. Since these limits recover the underlying critical point of the constant threshold system, we call this mechanism by Adaptive Near-Critical Coding (see Discussion).

For weak inputs, the entropy is maximized for $\tau = 1,000$ms, yielding a broad region around which PC can be efficiently performed (Figure 3A-B). Recall that the entropy, here, can be interpreted as the diversity of generated patterns. Conversely, the MI is maximized for a somewhat larger timescale, around $\tau = 1,100$ms, since the RC capacity keeps growing even when the entropy starts dropping. Again, these maxima are explained by the near-critical fluctuations $\rho \sim 1/\tau$. Both small and large $\tau$ yield only stereotyped activity, which decrease the entropy and leave a maximum for some intermediary time scale.

This information processing scenario is also robust with respect to synaptic coupling. Both the weak inputs' entropy (Figure 3D-left) and the MI (Figure 3D-right) maxima are persistent as $J$ is swept around the critical point. This is in a dramatic contrast with constant threshold networks, making adaptation an ideal and physiologically plausible mechanism for weak input processing via variance-encoding. These results are reproduced for alternative adaptation rules, although the entropy peak shifts to $\tau = 10^2$ms (Figure S8). Pyramidal CA3 and mossy neurons in the hippocampus (Trinh *et al.*, 2023), as well as neurons in the fish pallium (a hippocampus-like structure) (Trinh *et al.*, 2019), were reported having $\tau$ in the range of $10^2$–$10^3$ms.

For completeness, we show that PLs in the avalanche distributions are distorted by threshold adaptation. The distortion fades to the critical scaling of MF-DP within a limited range of avalanche sizes as $\tau$ is increased (Figure S7), as expected.

### Discussion

*A. A mechanism of robust optimal PC*

We compared rate and pattern coding capabilities of recurrent excitatory networks without and with threshold adaptation using the dynamic range, response entropy and input-output mutual information. The encoding of the adaptive network is robust against substantial changes in synaptic strengths. Both external and synaptic inputs contribute to increasing thresholds, although neural fatigue tends to stabilize network activity by reducing firing and self-suppressing noise.

# Network coding properties: adaptive threshold

## A. I/O phase diagram
## B. Coding capacities
## C. Coding scheme

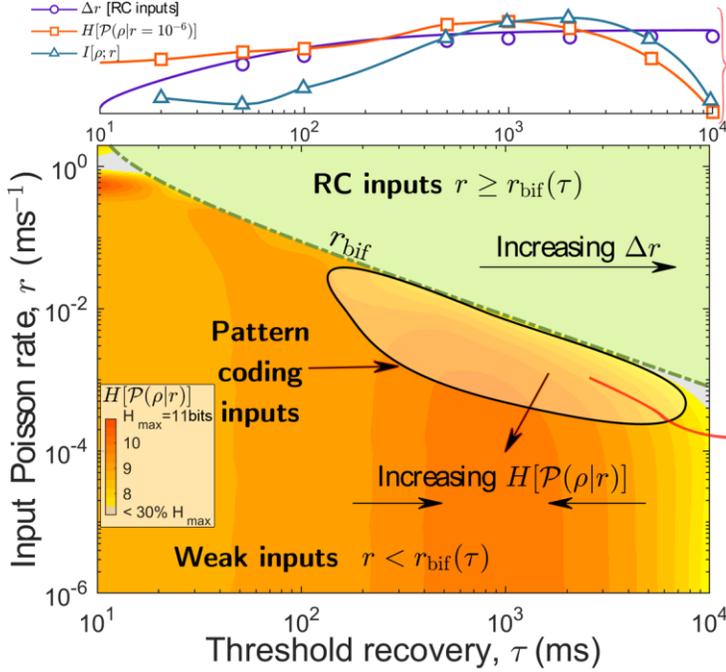
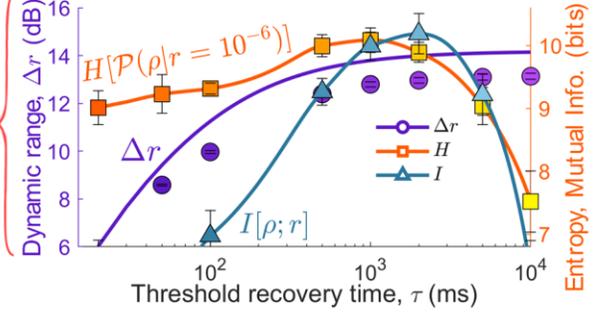
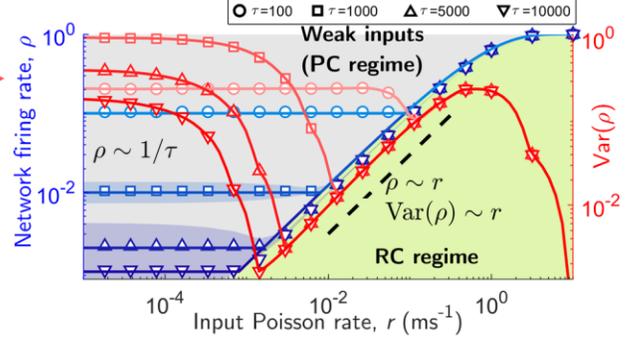

## D. Robust coding

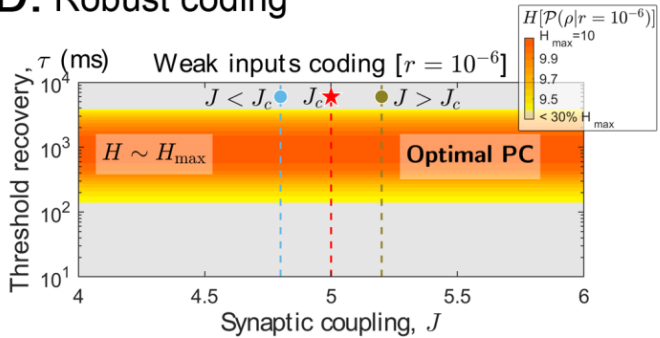
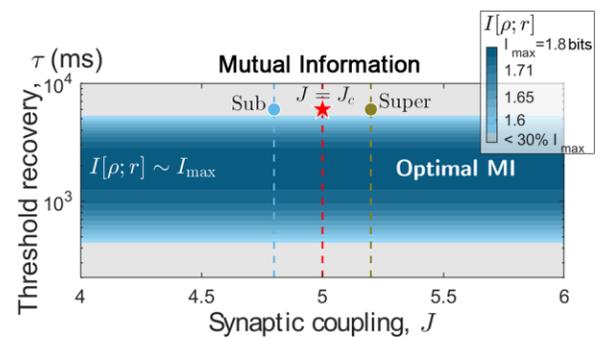

**Figure 3.** **Coding with adaptive threshold. A (top):** Coding capacities versus threshold recovery time: dynamic range=circles, entropy for weak inputs=squares ($r = 10^{-6}$ms$^{-1}$), and MI=triangles. **A (bottom):** I/O phase diagram extending the curves on the top panel: best coding strategy for each input depending on the recovery time $\tau$ (the diagrams in this figure have $J = 5$ but are reproduced for all considered $J$). Green region (height=$\Delta r$): RC regime; dot-dashed green curve: lower bound $r_{\text{bif}}$ for RC below which the inputs are weak. Yellow-orange (within highlighted region): PC regime where the entropy increases with decreasing $r$ (yellow=smaller and orange=larger entropy; gray=less than 30% of maximum $H$). **B:** Same as in A(top), but in full detail. MI and $H$ are optimized for $\tau \sim 10^3$ms, and $\Delta r$ grows with increasing $\tau$. The entropy peak is robust as a function of $J$ (Figure S5C) and shifts to $\tau = 10^2$ms for alternative threshold dynamics, being also a consequence of the variance of $\rho(t)$ (Figure S8C-S8D). Symbols=simulations (complete graph with $N = 10^5$ neurons; error bars are standard deviations). Solid curves=analytical results ($\Delta r$) or inter/extrapolation ($H$ and MI). See Supplementary Information for the calculation details. **C:** Stimulus-response (RC) curves (blue, mean of $\rho$) and $\text{Var}(\rho)$ (red) for any considered $J$ (subcritical, critical, supercritical; symbols=$\tau$ values); the stimulus-response curves flatten as r decreases preventing RC, but $\text{Var}(\rho)$ encodes $r$, allowing PC. **D:** Weak inputs entropy (left) and MI (right) are optimized within the considered range of $J$, highlighting the robustness of the system in contrast to constant threshold networks (in which optimizations happen only near $J_c$); dashed lines with symbols=reference $J$ to compare with Figure 2.

This produces synaptic strength invariance (see Supplementary Material), since the threshold $\theta$ enters as negative feedback in the firing rate $\rho$,

$$\rho(t+1) = [1-\rho(t)][I(r) + J\rho(t) - \theta(t)]\Gamma$$
$$\theta(t+1) = \theta(t) - \frac{\theta(t)}{\tau} + u\,F(\rho(t), \theta(t)) \qquad (1)$$

where $I(r) + J\rho$ are the external + synaptic inputs, $\Gamma$ is a constant gain, $0 < u \leq 1$ is the fatigue strength, $\tau > 1/u$ is a slow recovery timescale, and $F(\rho, \theta) = \rho\theta$ or $F(\rho, \theta) = \theta$ is the driving function. All parameters and variables are population averages. The constant threshold dynamic requires $\tau \to \infty$ and $u \to 0$, yielding the previously mentioned MF-DP critical point $J_c = 1/\Gamma = 5$ when $I(r) = \theta$.

Equation (1) has two solutions: a stationary state $\rho \sim 1/\tau$ for weak inputs ($r \to 0$), and $\rho \sim r$ when inputs are non-negligible. The stationary solution is the typical signature of self-organized quasicritical (SOqC) systems that hover around an underlying MF-DP critical point(Kinouchi et al., 2020). However, our system cannot be classified as SOqC since, contrary to the usual SOqC dynamics, here $J$ is a free parameter, so the critical point $J_c$ is not an attractor of Equation (1), and we consider arbitrary and simultaneous inputs. The threshold feedback alone is capable of regularizing network activity by weakening the dependence on $J - J_c$ to be of order $1/\tau$. Increasing $\tau \gg 1/u$ (with fixed $u \ll 1$) produces maximal dynamic range and avalanches that approach the MF-DP scaling in the weak inputs' regime. This would suggest that the RC regime of the adaptive net should also follow the MF-DP law where $m = 1/\delta_h = 1/2$; but this is not the case. The adaptive network processes RC inputs suboptimally, with $m = 1$, and hence has smaller dynamic range than its constant threshold counterpart. Notwithstanding, weak stimuli generate the stationary $\rho \sim 1/\tau$ firing rate with maximal variance for $\tau$ within $10^2$ to $10^3$ms, allowing for information to be reliably encoded into patterns of simultaneously active neurons. These timescales are observed in many circuits involved in higher brain function like memory (see below).

Current SOqC theory and its variants offer no insights about the coding strategies of neural systems, neither they consider the isolated effect of threshold adaptation. Here, we offered a comprehensive account of criticality and coding, and the dual representation enabled by threshold adaptation, calling it Adaptive Near-Criticality Coding – a cousin theory to SOqC.

### B. The role of adaptive threshold networks

We isolated the effect of firing threshold adaptation, and linked it to different forms of information processing: RC and PC. Many experimental and theoretical findings support the (near-)critical brain hypothesis, although there is no consensus about whether or not the brain operates through a self-organized (quasi-)critical process (Girardi-Schappo, 2021). It is reasonable to assume that the brain should maintain relatively small firing rates due to constraints in energy consumption and information processing flexibility (Denève & Machens, 2016). We showed that such self-controlled activity can be accomplished by threshold fatigue, making the coding reliable and robust against gross changes in synaptic strengths for relatively long recovery timescales. Threshold-adapting networks can process signals using two strategies: PC optimally transduces weak stimuli into variance-encoded spike patterns of collectively activated neurons, while RC captures inputs of larger intensity, yielding maximal MI. Future work could address feed-forward coupling of our adaptive populations, aiming at unveiling a relation between PC (finite-size fluctuations of $\rho$) and information filtering (Deger *et al.*, 2014).

At a basic level the negative feedback of $\theta$ onto $\rho$ resembles an effective inhibitory current since it reduces the mean network firing rate. This suggests that our results could be extended to include inhibitory dynamics known to be present in various motifs in biological neural nets. Such inhibition could help stabilizing the metastable states seen here for multiplicative adaptation. Adaptation allows decoding weak stimuli by the accumulation of past activity (Naud & Gerstner, 2012). Here we provided an alternative dual strategy for the optimal encoding of information when longer-lived adaptation is present.

*Threshold adaptation could add PC to the repertoire of sensory systems*. Adaptation is found in many neurons in isolation or in networks along sensory pathways (Wark *et al.*, 2009; Lundstrom *et al.*, 2010). It can perform temporal decorrelation (Wang *et al.*, 2003; Benda, 2021; Nesse *et al.*, 2021), enhance weak signal detection (Ratnam & Nelson, 2000; Chacron *et al.*, 2001) and act simultaneously on many timescales (Lundstrom *et al.*, 2010). Experimental evidence that codes need to be finer than just a simple accumulation of spikes (Jacobs *et al.*, 2009) support the requirement for variance-encoded spike patterns as an strategy. PC offers many advantages (Olshausen & Field, 2004), and there is evidence for it in the auditory (Feigin *et al.*, 2021), olfactory (Saha *et al.*, 2015) and visual (Osborne *et al.*, 2008; Jacobs *et al.*, 2009) systems.

*Adaptation could boost separation of similar patterns in memory systems*. Adaptive neurons occur at higher levels of neural processing, *e.g.* hippocampus, cortex and pallium in the context of memory formation. Hippocampal networks are required for identification (pattern separation) of the sparsely encoded input (Bakker *et al.*, 2008; Berron *et al.*, 2016; Knierim & Neunuebel, 2016). Hippocampal neurons have highly adaptive thresholds with long recovery times (Trinh *et al.*, 2023) and form recurrent networks that are essential for the storage and retrieval of specific memories (Hainmueller & Bartos, 2020). In particular, CA3 and hilar mossy cells have strong threshold adaptation with decay times of the order of $10^2$ to $10^3$ms (Trinh *et al.*, 2023). Similar timescales were found in the dorsolateral pallium of the teleost fish (Trinh *et al.*, 2019), a hippocampal-like region also engaged in memory of spatial patterns. Thus, we hypothesize that the PC capacity of these networks is due to their recurrent excitatory connectivity and the long-timescale of the spike threshold adaptation of its constituent neurons.

Beyond neuroscience, our findings have potential applications in artificial intelligence, sensors and cryptography. For example, *excitable networks of adapting elements could enhance reservoir computing*, serving as the reservoir component in neural networks. These networks rely on driving a reservoir of excitable elements with input patterns to generate a broad range of response waveforms. Through supervised learning and regression, these waveforms can then be combined to match inputs to unique output patterns, making them particularly suitable for sequential data processing (Tanaka *et al.*, 2019). The additional degree of freedom provided by the adaptation variable could further enhance high-dimensional sparsification (Cayco-Gajic *et al.*, 2017) by promoting PC, thereby improving the performance of reservoir computing. By adjusting timescales to optimize MI via a PC-RC dual coding strategy, we anticipate performance gains greater than or equal to those achieved with more complex adaptive schemes involving both threshold and synaptic plasticity (Lazar *et al.*, 2009).

*Adaptive networks could also improve the design of artificial sensors* (López-Higuera, 2021), where the network's optimal PC capacity could enhance detection sensitivity. Such sensors would be capable of detecting both weak and strong signals using different modes for each range. Additionally, maximal pattern variability in adaptive networks may have applications in cryptography, where they could function as code scramblers. A challenge in this context would be to reliably generate and reverse the patterns to recover the original message (Christof & Pelzl, 2009). Finally, the simplicity of our adaptive network makes it suitable for implementation in programmable neuromorphic chips, facilitating its use in these areas.

PC is not new. However, here we provided a mechanistic explanation for how the amplification of this coding capacity emerges from the collective response of excitable units with firing threshold adaptation. Such units are found within recurrent networks throughout the brain and are implicated in many functions, from sensory systems to memory, and could support the emergence of brain oscillations (Lombardi *et al.*, 2023). Even though we studied the system in the context of the brain, the model is a simple and general excitable bivariate stochastic process under the external influence of random signals that can provide clues for adaptive collective behavior beyond physics and neuroscience.

## Methods

### A. The stochastic model

We chose to study a general stochastic model that has been used to replicate experimental avalanche data (see Supplementary Material for a quick review). We study a network of $N$ excitatory neurons modeled by stochastic leaky integrate-and-fire units with discrete time step equal to 1ms. The network is an all-to-all graph because we want to isolate the threshold adaptation phenomenon, regardless of network architecture; in turn, this allows the development of a mean-field theory to describe the mean firing rate of the system. A binary variable indicates if the neuron $i$ fired at time $t$ [$X_i(t) = 1$] or not [$X_i(t) = 0$]. The membrane potential of each cell evolves as:

$$V_i(t+1) = \left[ I_i(t) + \mu_i V_i(t) + \frac{1}{N}\sum_{j \neq i}^N J_{ij} X_j(t) \right] [1 - X_i(t)] \quad (2)$$

where the $0 \leq \mu_i < 1$ are the leakage inverse time constants, and $I_i(t)$ models a constant input or membrane bias to neuron $i$. The synaptic parameters are $J_{ij} > 0$ (excitatory coupling strengths). The term $1 - X_i(t)$ automatically implements the reset to $V_i(t+1) = 0$ when $X_i(t) = 1$. This makes the membrane potentials strictly positive. The firing variable takes the value $X_i(t+1) = 1$ according to the conditional probability $\Phi(V_i(t)) \equiv \mathcal{P}(X_i(t+1) = 1|V_i(t))$, a piecewise linear sigmoidal transfer function of the form:

$$\Phi(V) = \begin{cases} 0 & V < \theta_i \\ (V - \theta_i)\Gamma_i & \theta_i \leq V \leq V_i^{(s)} \\ 1 & V > V_i^{(s)} \end{cases} \quad (3)$$

where the $\theta_i$ are the firing thresholds, $\Gamma_i$ are the firing gain constants (*i.e.*, the slope of the firing function), and

$$V_i^{(s)} = \theta_i + \frac{1}{\Gamma_i} \quad (4)$$

are the saturation potentials ensuring the continuity of the function when $\Phi(V) = 1$. An alternative way to include the input is by modifying the firing function $\Phi(V)$ to include a Poisson term (Methods Section C). Notice that this model intrinsically generates threshold variability, since a neuron can fire at any potential $V_i > \theta_i$ with probability $\Phi(V_i)$. Also, an isolated neuron ($J_{ij} = 0$) spontaneously spikes if the bias $I_i$ is chosen to be larger than the threshold $\theta_i$, since then $V_i(t) \geq I_i > \theta_i$, yielding the firing probability $\Phi(V_i(t)) > 0$ for any time $t$.

Intuitively, $\Phi(V)$ in equation (3) is the probability of each cell emitting a spike given it has membrane potential $V$; a larger membrane potential implies a larger chance of spiking. This probability is zero for $V < \theta_i$ (defining the absolute threshold, $\theta_i$); then, it grows linearly between $V = \theta_i$ and $V = V_i^{(s)}$ with slope $\Gamma_i$ (in this regime, the cell has a nonzero chance of emitting a spike, but the firing is stochastic); and finally, it saturates for $V > V_i^{(s)}$, generating deterministic firing.

Note that the transition to $X_i(t) = 1$ in step $t$ is always followed by $X_i(t+1) = 0$. This is because at $t + 1$, $V_i(t+1) = 0$ and $\Phi(0) = 0$, resulting in an absolute refractory period of 1ms.

Thus, when $V > V_i^{(s)}$, the spikes are deterministic and occurring with period 2ms. The $X_i$ variables are independently evaluated according to $\Phi(V_i(t))$ for each neuron $i$ at each time step $t$.

## B. Threshold adaptation

Many neurons in the brain present a dynamic firing threshold, from hippocampal neurons involved in navigation and learning (Trinh *et al.*, 2019; Trinh *et al.*, 2023) to neurons that participate in sensory perception (Pratt & Aizenman, 2007). We implement a very simple type of activity-dependent plasticity of intrinsic excitability by letting the threshold respond to each spike,

$$\theta_i(t+1) = \theta_i(t) - \frac{\theta_i(t)}{\tau_i} + u_i \theta_i(t) X_i(t) . \tag{5}$$

The parameters $0 \leq u_i \leq 1$ are the fractional increase of spike threshold at every firing (or spike fatigue; a large $u_i$ means a greater resistance to high firing activity), and the $\tau_i > 1/u_i$ are the recovery time scales of the adaptation mechanism. This dynamic enters the model in equation (3), shifting the entire curve of $\Phi(V)$ toward larger $V$ and effectively blocking the spiking of neurons with $V_i(t) < \theta_i(t)$.

This type of firing threshold dynamics is known to generate spike frequency adaptation by mimicking the slow inactivation of Na⁺ channels accumulated by successive spikes (Benda *et al.*, 2010; Platkiewicz & Brette, 2011). Notice that the larger the threshold, the larger the increment it gets from a spike, since the driving term is $u_i \theta_i X_i$ characterizing a multiplicative effect of $X_i$. This is because a neuron that is fatigued from spiking frequently (*i.e.*, having large $\theta_i$) should become proportionally more fatigued than a neuron that is not spiking, mimicking the depletion of threshold-related ions, and the saturation of the corresponding ionic channels. We tested two alternative driving mechanisms for the threshold (*i.e.*, an additive effect of X and a saturating multiplicative effect). All the adaptive rules produced the same qualitative results. See the Supplementary Material for full details.

When there is no spike, the threshold decays towards $\theta_i(t) \to 0$ following equation (5). However, remember that the neuron spontaneously spikes whenever $\theta_i(t) < I_i$. For this reason, the neuron resumes firing when $\theta_i(t)$ falls below the bias $I_i$, making the threshold grow again. $I_i$ is thus a lower bound for the threshold. In other words, equation (5) implements a homeostatic threshold around $I_i$.

## C. External stimulus

We chose the input to any given neuron to be a Poisson process generating an average of $r$ spikes every millisecond. This process acts homogeneously and independently on every neuron of the system, mimicking random input patters coming in that must be processed by this network. This implies that each neuron in the network will fire due to external stimulation with probability per unit time

$$P(r) = 1 - e^{-r}. \tag{6}$$

This probability is constant for a fixed input rate $r$.

This is implemented in simulations by taking advantage of the stochastic nature of the model. We redefine the firing function to reflect the fact that this external Poisson input can also cause a neuron to fire with probability $P(r)$:

$$\Phi_{total}(V) = \Phi(V) + P(r) - \Phi(V)P(r), \quad (7)$$

where $\Phi(V)$ is given by equation (3). In this case, a neuron $i$ will emit a spike with conditional probability $\mathcal{P}(X_i = 1|V_i) \equiv \Phi_{total}(V_i)$ instead of $\Phi(V_i)$ alone. Equation (7) takes into account the coincidence of externally and internally generated spikes by subtracting the intersection of both events.

## D. Simulation and Measurements

Equations (2) to (4) define the *constant threshold* model, since the $\theta_i$ are regarded as constant parameters. Conversely, the set of equations (2) to (5) is called the *adaptive threshold* model. In both cases, the input bias is kept constant at $I_i = 1$ for all neurons. For the constant threshold model, we then assume $\theta_i = 1$, such that any spike in neuron $i$ is either a consequence of the input Poisson process, or was generated from the sum of synaptic inputs. If we took $\theta_i < I_i$, neurons would spontaneously fire even when $r = 0$, causing an extra bias and confounding the Poisson input. All the details about this choice are discussed in the Supplementary Material. For the adaptive threshold model, we use $\theta_i(0) = 1$ as the initial condition.

In both versions of the model, the external stimuli are implemented via the firing function in equation (7). All parameters are assumed to be self-averaging (*i.e.*, they independently obey single-peaked distributions with small coefficients of variation, such as sharp Gaussians), yielding well-defined averages over the network elements for all quantities. The average parameters of the network are given by $\mu = \langle \mu_i \rangle$, $\Gamma = \langle \Gamma_i \rangle$, $\tau = \langle \tau_i \rangle$, and $u = \langle u_i \rangle$, where the angle brackets $\langle . \rangle$ represent the population average. The average interaction is $J = \langle J_{ij} \rangle = (1/N^2)\sum_{i \neq j} J_{ij}$. In summary, symbols that have no subscripts represent their average network value. The average leak parameter is $\mu = 0$ for both models, since this does not affect the universality class of the phase transition (Girardi-Schappo *et al.*, 2021), see below. In the adaptive threshold model, we fix the average threshold increase $u = 0.1$. Both $u$ and $\mu$ are homogeneous over the network.

The constant threshold model has a phase diagram with $I + h - \theta$ in the vertical axis and $J$ in the horizontal one (Figure S1). The parameter $h$ can be seen as an effective external field, such that at $h = 0$ (with $I = \theta$) there is a mean-field directed-percolation (absorbing) phase transition for mean synaptic coupling $J = J_c = 1/\Gamma$ (Girardi-Schappo *et al.*, 2021). The region with average coupling $J < J_c$ is subcritical (exponentially dying activity), and $J > J_c$ is supercritical (self-sustained branching process activity). The critical point $J_c = 1/\Gamma$ presents activity with power-law distributed avalanches, as expected (Figure S1). Injecting Poisson spikes with $r > 0$ is equivalent to taking $h > 0$, so we overlay the I/O phase diagram from Figure 2 onto the $(I + h - \theta) \times J$ plane. For both models, we fix $\Gamma = 0.2$ and use $J$ as a control parameter around the critical point $J_c = 1/\Gamma = 5$. The other control parameter is $r$ (the external input rate used to probe for the dynamic range), allowing the exploration of the $(I + h - \theta) \times J$ plane in its entirety. For the adaptive model, $\tau$ is also used as a control parameter for the recovery timescale of $\theta(t)$.

The quantities $r$ and $\rho$ are, respectively, input and output spike firing rates; *i.e.*, number of spikes per unit time per neuron. In a network with $N$ neurons, the instantaneous output firing rate is computed as

$$\rho(t) = \langle X_i(t) \rangle = \frac{1}{N} \sum_{i=1}^{N} X_i(t) \tag{8}$$

where $X_i(t)$ is the spiking variable. Equation (8) is called the population-averaged firing rate and is time-dependent. Simulating the network for a fixed input $r$, we use the fluctuations of $\rho(t)$ to empirically compute the distribution of patterns $\mathcal{P}(\rho|r)$, which is then used to compute the conditional Gibbs-Shannon entropy, $H[\mathcal{P}(\rho|r)]$, see Methods Section F.

The time-averaged output firing rate for a given $r$ is

$$\begin{aligned}\rho &= \frac{1}{T} \sum_{t=t_0}^{t_0+T} \rho(t) \\ \theta &= \frac{1}{T} \sum_{t=t_0}^{t_0+T} \langle \theta_i(t) \rangle\end{aligned} \tag{9}$$

where $t_0$ is some convenient time, and $T$ is a time interval. We employ equation (9) to characterize the rate coding (RC) regime (*i.e.*, the stimulus-response curves and the dynamic range are obtained by simply plotting $\rho$ vs. $r$). To be able to measure the fluctuations of $\rho$ in the simulation of networks with $r = 0$, we use the standard infinitely slow driving (Buendía *et al.*, 2020), in which a neuron is perturbed at the end of each avalanche, immediately generating a subsequent firing cascade.

Measurements of the temporal averages $\theta$ (when the threshold is dynamic) and $\rho$ are made over the steady state values of the variables $\rho(t)$ (for the constant threshold), or $\rho(t)$ and $\theta(t)$ (for adaptive threshold). The only exception is the measurement for high input rate $r$ in the adaptive threshold model. In this case, the driving term $\langle u_i \theta_i(t) X_i(t) \rangle = u\theta(t)\rho(t)$ generates an exponentially increasing threshold for strong inputs (see Supplementary Material). After a transient time $t = D$, the threshold reaches a sufficiently large value and the activity of the network completely shuts down, $\rho(t \geq D) = 0$. In other words, strong inputs cause the stable network activity to be $\rho = 0$ in the long time, although the network output rate responds to the input during a transient time $D$, allowing RC via the transient metastable state in which $\rho(t < D) > 0$. The lower bound $D \sim 1/u$ is calculated in the Supplementary Material, and is shown to be much larger than that in simulations in Figure S5B. Thus, in the adaptive model, the temporal averages $\rho$ and $\theta$ in equation (9) are taken over the transient metastable state, such that $t_0 = 0$ (the instant in which the Poisson input is turned on) and $T = D$ (the instant just before the activity shuts down). Both the alternative threshold driving rules present fully stable RC regimes ($D \to \infty$). All the details are given in the Supplementary Material.

We also measure the avalanche distributions of the adaptive model by thresholding the $\rho(t)$ timeseries (Figure S7). All the details are given in the Supplementary Material. Since our model without adaptation has a known critical point, we studied the effect of both the adaptive mechanism and the Poisson input on the shape of the distributions and their related crackling noise scaling relation. It is important to emphasize that we are not using the avalanche-related metrics to claim the existence of the critical point, as usually done, since the existence of the critical point is guaranteed by the theory for the constant threshold model.

## E. Dynamic range and Rate Coding quantification

The firing rate response of the network, $\rho$, as a function of the input rate $r$, has a minimum $\rho_{min} = \min_r \rho(r)$ and a maximum $\rho_{max} = \max_r \rho(r)$, rising smoothly in between these extremes. The dynamic range $\Delta r$ is the logarithmic range of $r$ within which there is an observable change in the level of the output $\rho$. We call $r_{10\%}$ the value of $r$ where the output level is 10% above its minimum. We also need the value of the input where the output level is 10% below the maximum, that is $r_{90\%}$, and we define (illustration in Figure 1C):

$$\Delta r = 10 \log_{10} \left( \frac{r_{90\%}}{r_{10\%}} \right). \tag{10}$$

We assume that no reliable rate coding can occur below $r_{10\%}$ or above $r_{90\%}$. Within these bounds, rate coding is characterized by the Stevens exponent $m$ (Stevens, 1957),

$$\rho \sim r^m. \tag{11}$$

For mean-field branching processes, the Stevens exponent can be shown to be $m = 1/\delta_h$ at the critical point, where $\delta_h$ is the external field critical exponent (Kinouchi & Copelli, 2006) – see the derivation for our model in the Supplementary Information. This law defines the RC regime. A given input is *weak* if it cannot be reliably encoded into output rates, typically having $r$ on the order of $10^{-6}$ spikes/ms (that is, an average of 1 spike every 1,000s arriving in each neuron) or smaller.

We employ the dynamic range $\Delta r$ to quantify the capacity of the system to perform RC. Theoretical analysis has shown that $\Delta r$ is optimized at the critical point (Kinouchi & Copelli, 2006), and this was verified for our model and experimentally (Shew *et al.*, 2009). Smaller $m$ yields larger dynamic range and better sensitivity to weaker inputs via RC. In other words, larger dynamic ranges extend further to smaller input intensities (*i.e.* Poisson rates $r$), making the system more susceptible to discern between signals with small rates. Thus, larger dynamic range implies a better capacity for RC.

In practice, we determine the level of the output firing rate at a percentage $\alpha$,

$$\rho_\alpha = \rho_{min} + \alpha (\rho_{max} - \rho_{min}), \tag{12}$$

where $\alpha = 0.1$ for the 10% output level and $\alpha = 0.9$ for the 90% level. Now, we need to find the value of $r = r_{10\%}$ that corresponds to $\rho_{0.1}$, and the input $r = r_{90\%}$ that corresponds to $\rho_{0.9}$. We have an analytic curve $\rho(r)$ for all the parameter combinations for both versions of the model (adaptive or not). Thus, we simply invert this expression to obtain $r_{10\%} = r(\rho_{0.1})$ and $r_{90\%} = r(\rho_{0.9})$ (see Supplementary Information for detailed calculations). In the simulations, we have a finite sample of $r$ and $\rho$ values. We then interpolate these data using MATLAB® Akima's cubic spline (which is best for sigmoid functions) in order to obtain $r(\rho)$. The smaller the input rate, the larger the time we need to run the simulation in order to get reliable results (simulation time is of order $1/r$). Therefore, we use $r = 10^{-6}$ ms$^{-1}$ as the reference input rate to determine $\rho_{min}$ to avoid running simulations for computationally prohibitive durations.

## F. Pattern Coding definition

In each and every instant, there is an average fraction $\rho N$ of neurons emitting a spike. These (distinguishable) spikes can be distributed over the $N$ neurons in $\Omega = N!/[(\rho N)!(N-\rho N)!]$ ways. Since the neurons are arranged in an all-to-all network, all the $\Omega$ patterns are equivalent (this is especially true in our mean-field approximation where the individual parameters are replaced by their respective population means). Thus, we only consider as distinct those patterns generated by different values of $\rho$, such that the fluctuations of $\rho(t)$ for a given $r$ dictate the capacity of the network to generate patterns. Hence, we say that the patterns are encoded in the finite-size variance of $\rho(t)$ (see Figure S6).

The entropy can be used to quantify the information in $\rho(t)$ for some $r$ and is then defined as the pattern coding (PC) capacity as follows. For a single fixed $r$, we estimate the distribution $\mathcal{P}(\rho|r)$ from the histogram of all the observed values of $\rho(t)$ during the stationary state. In the case of the network with multiplicative adaptation, this distribution is estimated during the metastable transient state. The entropy conditioned on a particular stimulus of constant rate $r$ is defined through the entropy functional, $H[\mathcal{P}] \equiv -\sum \mathcal{P} \log_2 \mathcal{P}$:

$$H[\mathcal{P}(\rho|r)] = -\sum_{\{\rho\}} \mathcal{P}(\rho|r) \log_2 \mathcal{P}(\rho|r), \qquad (13)$$

where the sum is taken over the set of all the observed outputs, $O = \{\rho\}$. Notice that this is the entropy conditioned on a single input $r$, whereas the conditional entropy is the average of Equation (13) over(14) all the inputs (see below).

## G. Mutual information

The set of inputs $S = \{r\}$ comprises all the input values that were used to stimulate the network for a fixed set of parameters. Since $r$ is a control parameter, $r$ is uniformly sampled within the interval of interest, making the set $S$ an ordered list containing $R$ equally spaced values. The Mutual Information is defined as $I[O; S] = H(O) - H(O|S)$. However, we can define a related information measure for systems that present spontaneous activity in the absence of input (Chacron *et al.*, 2007), i.e., in which $\mathcal{P}(\rho|r = 0) > 0$:

$$I[\rho; r] = H[\mathcal{P}(\rho|r = 0)] - \sum_{\{r\}} \mathcal{P}_R(r) H[\mathcal{P}(\rho|r)], \qquad (14)$$

where the second term is the conditional entropy $H(O|S)$ [an average of the entropy functional over the input set], and $\mathcal{P}_R(r) = 1/R$ since the inputs are uniformly sampled. This metric captures the drop in information due to inputs with respect to the ground level ($r = 0$), and is as reliable as the original definition (Chacron *et al.*, 2007).